\documentclass[showpacs,floatfix,citeautoscript,twocolumn,prb,superscriptaddress,nobibnotes]{revtex4-1}

\usepackage{graphicx,xcolor}
\usepackage{amsmath}
\usepackage{comment}
\usepackage{textcomp}
\usepackage{inputenc}
\usepackage{xspace}

\newcommand*{\aattt}{Au$_{1.9}$Sb$_{0.46}$Te$_{2.64}$\xspace}
\newcommand*{\C}{~$^{\circ}$C\xspace}
\begin{document}

\title{Synthesis of layered gold tellurides AuSbTe and Au$_2$Te$_3$ and their semiconducting and metallic behavior}

\author{Emma A. Pappas}
\affiliation{Department of Physics, University of Illinois at Urbana-Champaign, Urbana, IL, 61801, USA}
\affiliation{Materials Research Laboratory, University of Illinois at Urbana-Champaign, Urbana, IL, 61801, USA}

\author{Rong Zhang}
\affiliation{Stanford Institute for Materials and Energy Sciences, SLAC National Accelerator Laboratory, Menlo Park, CA, 94025, USA}
\affiliation{Department of Applied Physics, Stanford University, Stanford, CA, 94305, USA}

\author{Cheng Peng}
\affiliation{Stanford Institute for Materials and Energy Sciences, SLAC National Accelerator Laboratory, Menlo Park, CA, 94025, USA}

\author{Robert T. Busch}
\author{Jian-Min Zuo}
\affiliation{Materials Research Laboratory, University of Illinois at Urbana-Champaign, Urbana, IL, 61801, USA}
\affiliation{Department of Materials Science and Engineering, University of Illinois at Urbana-Champaign, Urbana, IL, 61801, USA}

\author{Thomas P. Devereaux}
\affiliation{Stanford Institute for Materials and Energy Sciences, SLAC National Accelerator Laboratory, Menlo Park, CA, 94025, USA}
\affiliation{Department of Materials Science and Engineering, Stanford University, Stanford, CA, 94305, USA}

\author{Daniel P. Shoemaker}\email{dpshoema@illinois.edu}
\affiliation{Department of Physics, University of Illinois at Urbana-Champaign, Urbana, IL, 61801, USA}
\affiliation{Materials Research Laboratory, University of Illinois at Urbana-Champaign, Urbana, IL, 61801, USA}
\affiliation{Department of Materials Science and Engineering, University of Illinois at Urbana-Champaign, Urbana, IL, 61801, USA}

\begin{abstract}
\centerline{\textbf{Abstract}}
Previous studies on natural samples of pampaloite (AuSbTe) revealed the crystal structure of a potentially cleavable and/or exfoliable material, while studies on natural and synthetic montbrayite (Sb-containing Au$_2$Te$_3$) claimed various chemical compositions for this low symmetry compound. Few investigations of synthetic samples have been reported for both materials, leaving much of their chemical, thermal and electronic characteristics unknown. Here, we investigate the stability, electronic properties and synthesis of the gold antimony tellurides AuSbTe and \aattt (montbrayite). Differential thermal analysis and \textit{in situ} powder x-ray diffraction revealed that AuSbTe is incongruently melting, while \aattt is congruently melting. Calculations of the band structures and four-point resistivity measurements showed that AuSbTe is a semiconductor and \aattt a metal. Various synthesis attempts confirmed the limited stable chemical composition of \aattt, identified successful methods to synthesize both compounds, and highlighted the challenges associated with single crystal synthesis of AuSbTe. 

\end{abstract}

\maketitle 

\section{Introduction} 

Heavy elements such as gold and tellurium have the potential to give rise to compounds with large spin-orbit coupling and narrow band gaps, but gold and tellurium-containing materials have a wide variety of properties that can be difficult to predict from their chemical formula and stoichiometry. As a result, the synthesis, band structures and properties of many gold tellurides are still unknown. The interesting characteristics and applications of certain materials in this family and of similar compounds justify further exploration of the gold tellurides. For example, calaverite AuTe$_2$ is a rare case of a natural mineral with an incommensurate crystal structure and is superconducting at high pressures;\cite{Kudo2013, Streltsov2018} small band gap semiconductors Bi$_2$Te$_3$ and Sb$_2$Te$_3$ are exfoliable topological insulators and useful thermoelectric materials;\cite{Shahil2012, Hsieh2009, Zheng2022} and the semiconducting van der Waals compounds $\alpha$- and $\beta$-As$_2$Te$_3$ have useful thermoelectric properties that can be improved upon substitution of As for Sn or Bi.\cite{Morin2015, Vaney2016} By investigating similar compounds in the Au--Sb--Te phase space, we are identifying materials with potentially interesting properties and applications. Here, we focus our efforts on the layered gold antimony tellurides AuSbTe and \aattt, the only ternary compounds in this phase space. 

During their respective studies of the Au--Sb--Te phase diagram, Blachnik and Gather (1976),\cite{BlachnikGather1976, GatherBlachnik1976} and Nakamura and Ikeda (2002)\cite{Nakamura2002} had synthesized AuSbTe years before the 2018 discovery of a natural specimen.\cite{Vymazalova2018} Blachnik and Gather reported that AuSbTe decomposes into Sb$_2$Te$_3$ and a liquid at 424\C and presented a diffraction pattern\cite{BlachnikGather1976}, but the crystal structure was only solved as part of a more recent study following the discovery of a natural sample. Pampaloite (AuSbTe), first found in the Pampalo gold mine in Finland, was described by Vymazalov\'{a} \textit{et al.} (2018).\cite{Vymazalova2018} The compound crystalizes in the C2/c space group. Its layered structure and chemistry makes AuSbTe a potential cleavable and/or exfoliable semiconductor, but these properties remain to be explored as very few investigations of natural and synthetic samples of pampaloite exist. Recent studies reported reflectance measurements, electron microprobe analyses, electron backscatter diffraction and Raman spectroscopy.\cite{Vymazalova2018, V2022} Here, we present synthesis, band structure calculations, and electrical resistivity measurements of AuSbTe. We also further investigate its stability through \textit{in situ} powder x-ray diffraction (PXRD) and differential thermal analysis (DTA). 

The rare mineral montbrayite was first described in 1946 by Peacock and Thompson,\cite{PeacockThompson1946} as they studied a specimen from the Robb-Montbray mine, Québec, Canada. The mineral was given the approximate chemical formula Au$_2$Te$_3$, but fusion experiments at this stoichiometric composition only gave calaverite (AuTe$_2$) and gold.\cite{PeacockThompson1946, Markham1960, Cabri1965} In 1972, Bachechi showed that the compound could be stabilized by partial substitution of Au and Te for Bi, Pd and Sb, or Sb alone. She found two stable synthetic compounds, Au$_{1.89}$(Sb,Pb)$_{0.11}$Te$_{2.88}$Bi$_{0.12}$ and Au$_{1.89}$Sb$_{0.23}$Te$_{2.88}$, with the former having a melting temperature of 410\C.\cite{Bachechi1972} Unlike AuSbTe, the composition of the Au$_2$Te$_3$-like phase is not consistent across previous studies. In 1976, Blachnik and Gather synthesized a compound with composition Au$_{1.65}$Sb$_{0.125}$Te$_{3.225}$ that decomposes into AuTe$_2$ and liquid at 460\C and forms at 444\C by reaction of this melt and AuTe$_2$.\cite{BlachnikGather1976, GatherBlachnik1976} In 2002, Nakamura and Ikeda revealed another composition for montbrayite, namely \aattt, and estimated its solid solution field to be limited to within one atomic percent of this chemical formula.\cite{Nakamura2002} 

More recently, Bindi \textit{et al.} (2018) performed a chemical and structural study on a natural specimen of montbrayite.\cite{Bindi2018} They found the centrosymmetric space group P$\Bar{1}$ to be a better fit than the previously reported non-centrosymmetric structure P1\cite{Bachechi1971}. Many studies have been performed on natural specimens of montbrayite,\cite{PeacockThompson1946, Bachechi1971, Bachechi1972, Criddle1991, Friedl1992, Edenharter1999, Genkin1999, ShackletonSpry2003, Bindi2018, V2023} but like pampaloite, the band structure and electrical transport of the compound are unknown. Optical properties\cite{Criddle1991}, microscopy techniques\cite{V2023}, reflectance,\cite{Criddle1991, V2023} Raman\cite{V2023} and M{\"o}ssbauer\cite{Friedl1992} spectra have been reported. More investigations of natural samples are reviewed by Bindi \textit{et al.}\cite{Bindi2018}, which is of limited interest for this paper, as we focus on the synthetic analog. Only a few synthetic samples have been investigated,\cite{Bachechi1972, BlachnikGather1976, GatherBlachnik1976, Nakamura2002} with most focusing on the chemical composition and stability of montbrayite. Those few studies report stable chemical compositions in the Au--Sb--Te phase diagram that are conflicting: Au$_{1.89}$Sb$_{0.23}$Te$_{2.88}$, Au$_{1.65}$Sb$_{0.125}$Te$_{3.225}$ and \aattt. To address these inconsistencies and reveal more of its characteristics, we synthesize montbrayite compounds with a range of chemical compositions, investigate its properties using \textit{in situ} powder x-ray diffraction (PXRD) and differential thermal analysis (DTA), and present new band structure calculations and electrical resistivity data.

\section{Methods}

A polycrystalline sample of AuSbTe was synthesized by mixing gold (99.99\% trace metal basis, $<$45~$\mu$m), antimony (99.5\%), and tellurium (99.8\% trace metal basis, -200 mesh) powders in stoichiometric ratios and sealing them in an evacuated silica tube of 10~mm inner diameter. The ampoule was placed in a muffle furnace, heated to 600\C at 5\C/min, annealed for 1~h, cooled to 350\C over 2~h, and annealed for 180~h before the furnace was turned off and cooled naturally. The sample was ground, pressed into a pellet, sealed in an evacuated silica tube, heated to 350\C at 5\C/min, annealed for approximately 185~h, then water quenched. These steps were repeated thrice with annealing times of 82~h, 190~h, and 168~h. Re-grinding and re-annealing proved to be essential and increased the phase purity of the product with each step. The incompletely reacted Sb$_2$Te$_3$, AuSb$_2$ and Au phases decreased with each annealing cycle as increasing amounts of AuSbTe formed. 

The polycrystalline sample of \aattt used for \textit{in situ} powder x-ray diffraction (PXRD) and differential thermal analysis (DTA) measurements was synthesized by mixing powders of gold (99.99\%, -100 mesh), antimony (99.5\%), and tellurium (99.8\% trace metal basis, -200 mesh) in an atomic ratio of 1.9:0.46:2.64. The reactants were sealed in an evacuated silica tube of 10~mm inner diameter. The ampoule was placed in a muffle furnace, heated to 600\C at 5\C/min, and annealed for 1~h before water quenching. The resulting ingot was ground and pressed into a pellet, which separated into a few large pieces as it was sealed in an evacuated silica tube. The sample was heated to 350\C at 5\C/min and annealed for two weeks before water quenching. The resulting product was re-ground, pelletized, and re-annealed at 350\C for 10 days, which proved to be an optional step as the sample was essentially unchanged. Pellets of \aattt lacked physical integrity, so the polycrystalline sample used for four point resistivity measurements was instead heated to 600\C at 5\C/min, annealed for 1~h, cooled to 350\C over 2~h, and annealed for 180~h before the furnace was turned off and cooled naturally. This heating profile produced a solid ingot with grains too small ($\lesssim$~200~nm) to perform high quality scanning transmission electron microscopy (STEM) measurements. The STEM sample was heated to 600\C at 5\C/min, annealed for 30~min, cooled to 450\C over 2~h, cooled to 350\C over 50~h, then cooled to room temperature over 2~h. This heating profile provided a sample that appeared to have multiple phases (dark spots and gold veins on a silver ingot), but the main phase \aattt had larger grains ($\lesssim$~1~$\mu$m) that were more appropriate for STEM imaging. 

Portions of the samples were crushed and analyzed using PXRD on a Bruker D8 diffractometer with a Mo x-ray source in transmission geometry. The crystal structures were refined using the Rietveld method with GSAS-II\cite{GSAS2}. For the \textit{in situ} PXRD measurements, the samples were diluted with ground quartz before being sealed in evacuated quartz capillaries. The heating rate was 60 \C/min. The fixed temperature scans lasted 93 minutes and were preceded by a 7-minute hold. 

Chemical composition of the samples was also verified using scanning electron microscopy (SEM) energy dispersive x-ray spectroscopy (EDS) analyses, the results and detailed description of which can be found in the Supporting Information.

Differential thermal analysis (DTA) measurements were performed on a Shimadzu Differential Thermal Analyzer DTA-50 with a heating rate of 20 \C/min. Crushed portions of the samples were sealed in evacuated tubes and Al$_2$O$_3$ powder was used as reference. 

Resistivity measurements were carried out using the 4-point contact probe method in a Quantum Design Physical Property Measurement System DynaCool. Using silver epoxy, four gold wire contacts were made on a rod of \aattt (1.46 $\times$ 1.05 $\times$ 5.84 mm$^3$). The rod was then mounted using Kapton tape. Silver epoxy and silver paint were used to make the four contacts on the AuSbTe rod (0.98 $\times$ 0.6 $\times$ 4 mm$^3$). The detailed description and results of the Van Der Pauw measurements on AuSbTe are in the Supporting Information.

The scanning transmission electron microscopy (STEM) sample was prepared by grinding the \aattt crystals into a fine powder in an agate mortar and pestle then dispersing onto an ultra-thin lacy carbon grid. The final samples were well dispersed particles several hundred nanometers in size. The STEM images were obtained using a FEI Themis Z (S)TEM operated at 300~kV. Atomic-resolution high-angle annular dark-field (HAADF) imaging was performed in STEM mode with a convergence angle of 18~mrad, a camera length of 115~mm, and a screen current of $\sim$29~pA with a 10~$\mu$s dwell time. EDS maps were obtained using a FEI Talos (S)TEM equipped with four windowless silicon drift detectors (SuperXTM) operated at 200~kV. EDS maps were collected with a 20.5~mrad convergence angle and a current of $\sim$85~pA. Data processing was performed in the on-board Velox software to calculate atomic percentages including only Au, Te, and Sb during atomic calculations.

The calculations of band structures using first principles method based on the density functional theory (DFT) were conducted with the VASP code\cite{Kresse1993,Kresse1994}, where VASP 5.4 PBE\cite{PBE} projector-augmented wave (PAW) pseudopotentials\cite{PAW1,PAW2} were used to carry out the calculations. The total valence electrons explicitly treated in our calculations of AuSbTe consist of Au s1d10, Te s2p4, and Sb s2p3, which are described by a plane-wave basis set with a cut-off energy of 550~eV. We performed the self-consistent calculations on the primitive cell aligned with the space group C2/c for bulk. The experimental lattice constants in primitive cell are $a$ = 11.947(3)~\r{A}, $b$ = 4.4812(10)~\r{A}, $c$ = 12.335(3)~\r{A}, $\alpha$ = 90$^{\circ}$, $\beta$ = 105.83(2)$^{\circ}$, and $\gamma$ = 90$^{\circ}$. In calculation, all atomic positions were relaxed using a conjugate gradient algorithm \cite{RevModPhys.64.1045}. The energy convergence tolerance was set at $1.0 \times 10^{-8}$~eV and the force was regarded as converged when smaller than $0.001$~eV~\r{A}$^{-1}$. The total valence electrons explicitly treated in the DFT calculations of Au$_2$Te$_3$ consist of Au s1d10 and Te s2p4, and with a cut-off energy of 500~eV. We performed the self-consistent calculations on the primitive cell aligned with the space group P$\Bar{1}$ for bulk. The experimental lattice constants in primitive cell are $a$ = 10.8045(7)~\r{A}, $b$ = 12.1470(7)~\r{A}, $c$ = 13.4480(7)~\r{A}, $\alpha$ = 108.091(5)$^{\circ}$, $\beta$ = 104.362(5)$^{\circ}$, and $\gamma$ = 97.471(5)$^{\circ}$, and the experimental structure is used in the calculations. Bands for both materials are plotted along the $k$-paths recommended by the SeeK-path tool\cite{HINUMA2017140}. 

The crystal structure drawings were produced with VESTA.\cite{Momma:db5098}

\textit{\textbf{Caution!} Incompletely reacted Sb or Te in sealed tubes can lead to high vapor pressures; synthesis conditions were chosen to avoid exposure to high temperatures when unreacted. }

\section{Results and Discussion}

\subsection{Structure of AuSbTe}

\begin{figure}
    \centering
    \begin{minipage}{\columnwidth}
        \centering
        (a)
        \includegraphics[width=0.42\columnwidth]{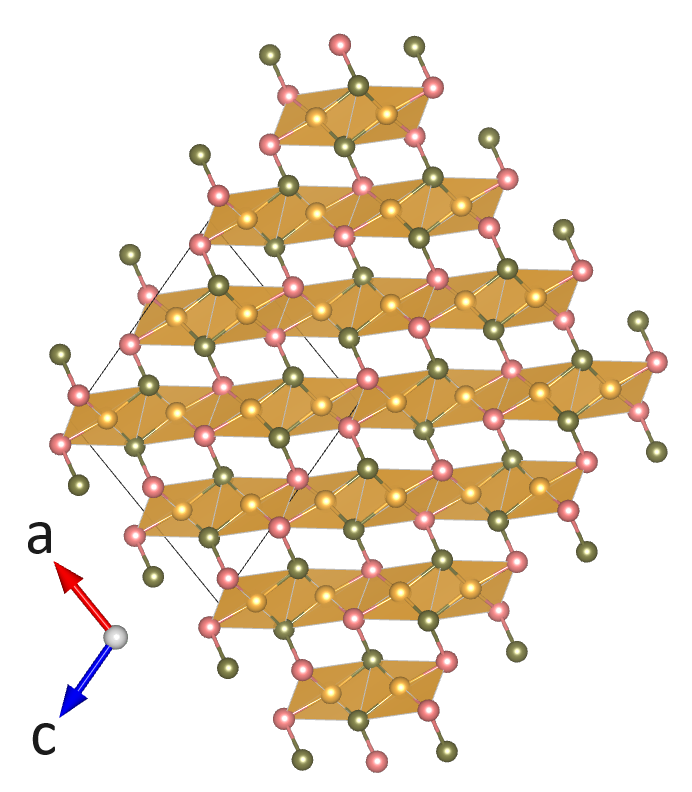}
        (b)
        \includegraphics[width=0.42\columnwidth]{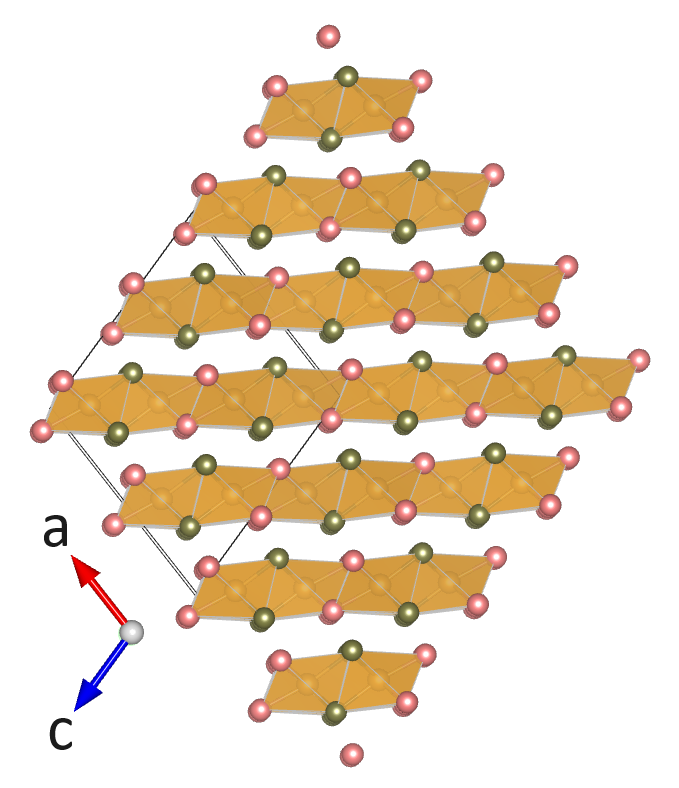}
    \end{minipage}
    \begin{minipage}{\columnwidth}
        \centering
        (c)
        \includegraphics[width=0.42\columnwidth]{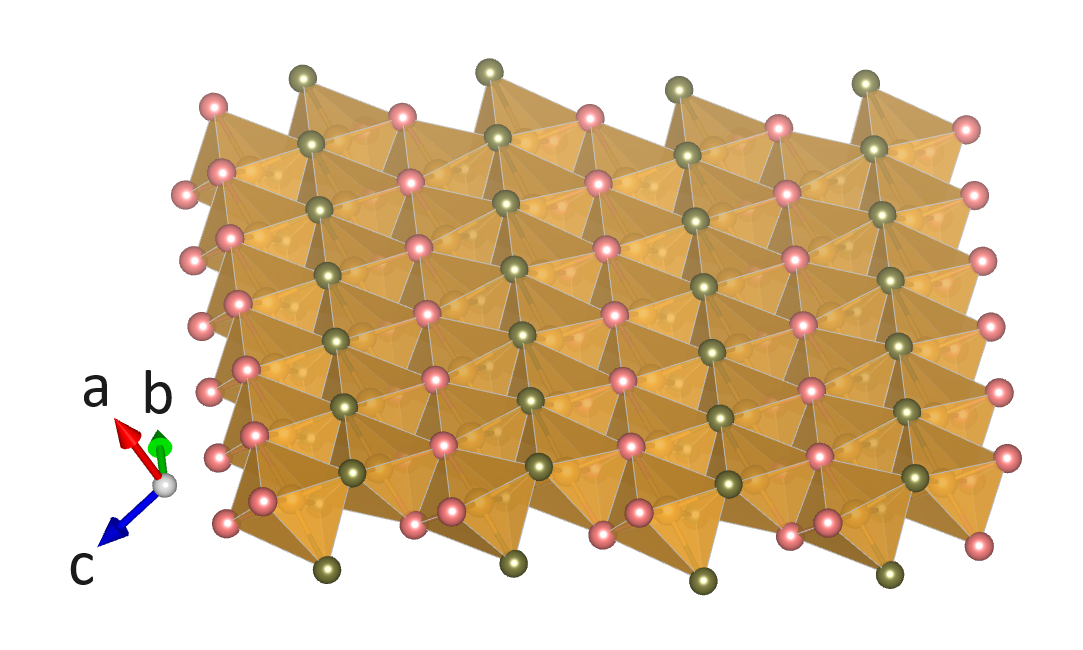}
        (d)
        \includegraphics[width=0.42\columnwidth]{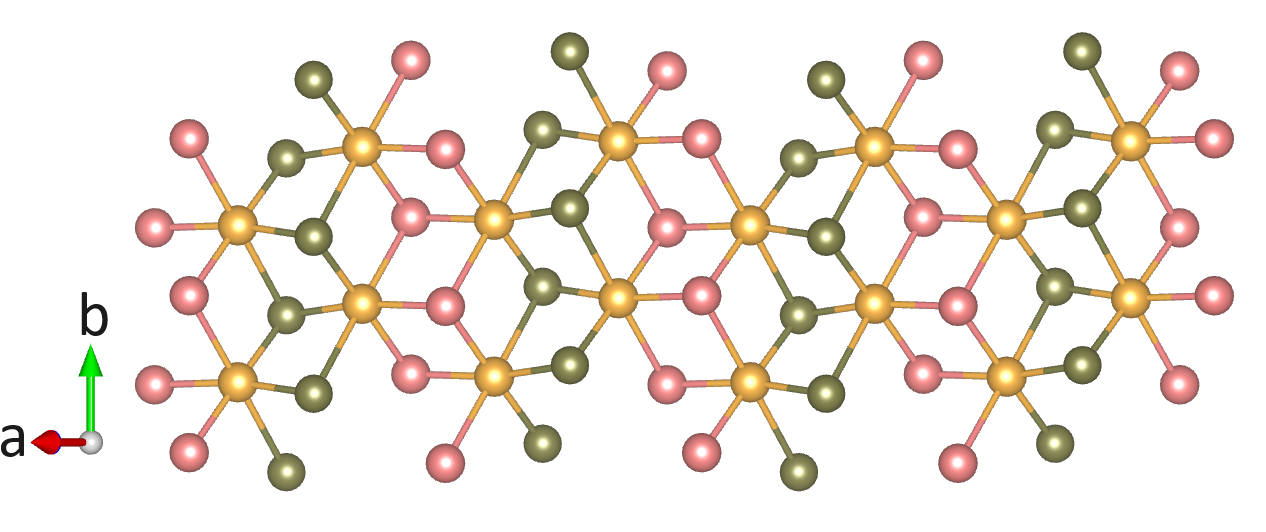}
    \end{minipage}
    \caption{Crystal structure of AuSbTe (C2/c). Au, Sb and Te atoms are shown in gold, pink and green. (a) Bonds shorter than 3~\r{A} are drawn to highlight the resemblance to the PdSe$_2$ verbeekite-type structure that has planar layers connected by bonds. (b)-(c) Some longer Au--Sb and Au--Te bonds are displayed instead of the interlayer Sb--Te bonds to show that Sb and Te form [AuSb$_3$Te$_3$] octahedra around Au, which create layers similar to the AuTe$_2$-type structure. (d) Top view of a single layer.  The atomic arrangement is akin to transition metal dichalcogenides in the distorted octahedral phase. }
    \label{fig:AuSbTe_struture}
\end{figure}

AuSbTe crystallizes in the monoclinic space group C2/c\cite{Vymazalova2018} and is very similar to both the PdSe$_2$ verbeekite-type structure (C2/c) and the AuTe$_2$-type structure (C2/m).\cite{Kempt2020} Fig.\ \ref{fig:AuSbTe_struture}(a) displays the atomic arrangement of AuSbTe in such a way to highlight its resemblance to verbeekite PdSe$_2$. The Au atoms (gold) are bonded to two Sb (pink) and two Te (green) atoms, with an additional bond between the Sb and Te atoms. The Au--Sb and Au--Te bonds range from 2.693(1) to 2.771(1)~\r{A}, while the Sb--Te bond is 2.829(2)~\r{A}.\cite{Vymazalova2018} In verbeekite, the Au--Te and Au--Sb bonds are replaced by four Pd--Se bonds of 2.4844(14) and 2.4966(11)~\r{A}, and the Sb--Te bond is replaced by a Se--Se bond of 2.442(2)~\r{A}\cite{Selb2017}. Fig.\ \ref{fig:AuSbTe_struture}(b)-(d) does not display the Sb--Te bond in AuSbTe and instead displays additional Au--Sb and Au--Te bonds of 3.071(1) and 3.257(1)~\r{A}.\cite{Vymazalova2018} With these longer bonds, the antimony and tellurium atoms form [AuSb$_3$Te$_3$] octahedra around the gold atoms, giving rise to flat single layers similar to those formed by the [AuTe$_6$] octahedra in the AuTe$_2$-type structure. In the latter, the Au--Te bonds (2.67 and 2.98~\r{A}) are all shorter than the Te--Te bonds ($\geq$3.20~\r{A})\cite{Schutte:bx0246, Reithmayer:se0099}, resulting in disconnected sheets of [AuTe$_6$] octahedra. This is in contrast to the actual structure of AuTe$_2$ (calaverite) where an incommensurate modulation creates Te--Te bonds shorter than some of the Au--Te bonds (2.88~\r{A}).\cite{Schutte:bx0246, scAuTe2} When seen from above, a single layer of AuSbTe (see Fig.\ \ref{fig:AuSbTe_struture}(d)) resembles that of transition metal dichalcogenides with distorted octahedral structure (TMDCs-1T). AuTe$_2$ also shares this resemblance with TMDCs-1T, but unlike the semiconducting AuSbTe (see the Electronic properties of AuSbTe section), it is a metal that becomes superconducting at high pressures or upon platinum doping.\cite{scAuTe2, Kudo2013} 

AuSbTe can be seen as replacing one Te$^{2-}$ ion from the AuTe$_2$-type structure by a Sb$^{3-}$ ion. Upon substitution with Sb$^{3-}$, the material can be charge balanced by forming Sb--Te dimers, effectively Au$^{3+}$(Sb--Te)$^{3-}$.
This Sb--Te bond is shorter (2.829(2)~\r{A}\cite{Vymazalova2018}) than the interlayer tellurium bond length in C2/m AuTe$_2$ (approximately 3.2~\r{A}\cite{Reithmayer:se0099}) and creates a resemblance to the structure of verbeekite. Interestingly, the verbeekite structure only appears after PdSe$_2$ undergoes a phase transition at high temperature and pressure\cite{Selb2017, Kempt2020} or when selenium is partially substituted with tellurium\cite{Wenhao2021}, and it has been predicted that, with more pressure, it could transform into the C2/m structure\cite{Lei2019, Kempt2020}. One could imagine that AuSbTe might also take the more layered C2/m structure at high pressure, which remains to be verified.  

\subsection{Stability of AuSbTe by synthesis, diffraction and calorimetry}

The particular C2/c layered structure of AuSbTe indicates its potential for being a cleavable and/or exfoliable material, which could be verified through the synthesis of large single crystals. Unfortunately, the presence of other stable competing phases in the Au--Sb--Te ternary phase diagram\cite{Prince1990, GatherBlachnik1976, ASMdiagram6732-4060} and its reported decomposition at 430\C (see the 67.1~at.\% Au, 32.9~at.\% Sb -- Sb$_2$Te$_3$ binary phase diagram\cite{Prince1990, ASMdiagram6732-4060}) makes single crystal syntheses of pampaloite difficult to achieve. In spite of these challenges, we have synthesized a polycrystalline sample  that makes the semiconducting nature of AuSbTe clear (see the Electronic properties of AuSbTe section). 

\begin{figure}
    \centering
    \includegraphics[width=\columnwidth]{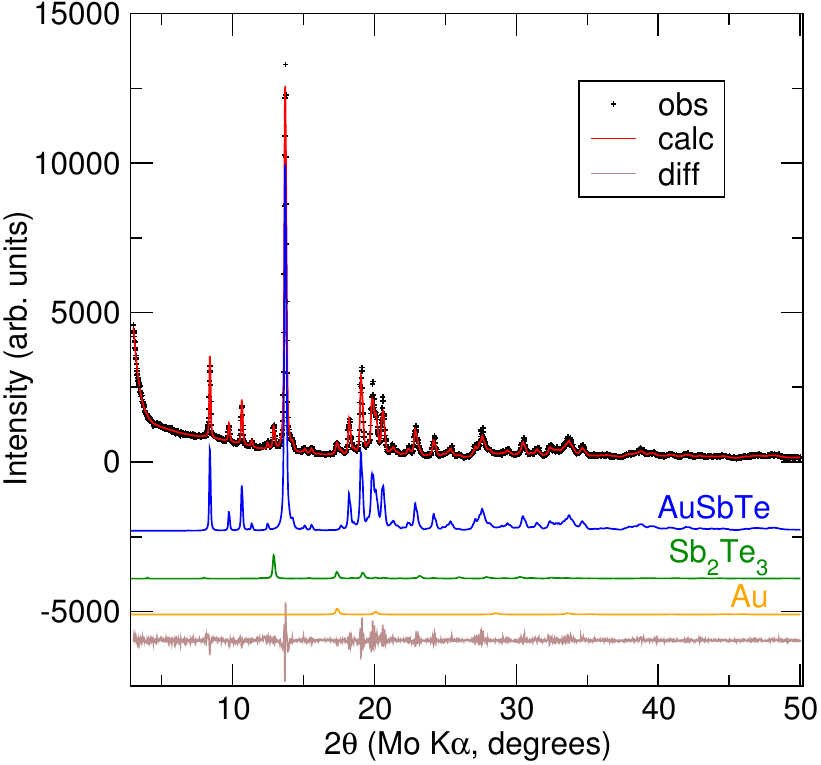}
    \caption{Rietveld refinement of the room temperature PXRD data of AuSbTe. Minor Au and Sb$_2$Te$_3$ impurities are present, corresponding to 1.2(2)~wt.\% and 6.8(3)~wt.\% respectively. }
    \label{fig:EAP0061_XRD}
\end{figure}

Among all of the synthesis techniques explored, the one resulting in the highest quality polycrystalline sample of AuSbTe is reported in the Methods section. It consists of grinding, pelleting, and annealing multiple times an ingot with stoichiometric composition. The success of this method is attributed to the repeated mixing and annealing, which increased the phase purity of the compound. Annealing was done at 350\C for a few days at a time. As can be seen in the PXRD analysis (see Fig.\ \ref{fig:EAP0061_XRD}) and in the SEM-EDS analysis (see Supporting Information), the sample has minimal (8~wt.\% combined) Au and Sb$_2$Te$_3$ impurities. Various portions of the as-synthesized pellet of AuSbTe were used for all measurements, so the PXRD and SEM-EDS results are representative of the chemical composition of all studied samples. The refinement shown in Fig.\ \ref{fig:EAP0061_XRD} was done with the atom positions of the AuSbTe phase fixed to the values from Vymazalov\'{a} \textit{et al.} (2018).\cite{Vymazalova2018} The lattice parameters of AuSbTe were included in the refinement, giving  $a$ = 11.956(3)~\r{A}, $b$ = 4.4827(6)~\r{A}, $c$ = 12.341(3)~\r{A}, $\alpha$ = 90$^{\circ}$, $\beta$ = 105.852(9)$^{\circ}$, and $\gamma$ = 90$^{\circ}$. The fit's overall weighted profile R-factor is 9.04\% and the goodness of fit is 2.19. 

Various synthesis attempts were made, with few giving the desired phase at all. Synthesis methods included diverse annealing temperatures, rates and hold times at stoichiometric composition; non-stoichiometric reactions based on the Au--Sb--Te phase diagrams\cite{Prince1990, GatherBlachnik1976, ASMdiagram6732-4060}; and chemical vapor transport with different transport agents. A recurrent finding was that Sb$_2$Te$_3$ is kinetically favored, making single crystal syntheses of AuSbTe difficult. No promising path for the slow cooling, the flux growth or the chemical vapor transport of single crystals of AuSbTe has been confirmed yet. The details of all of our synthesis attempts are available in the Supporting Information. 

\begin{figure}
    \centering
    \includegraphics[width=\columnwidth]{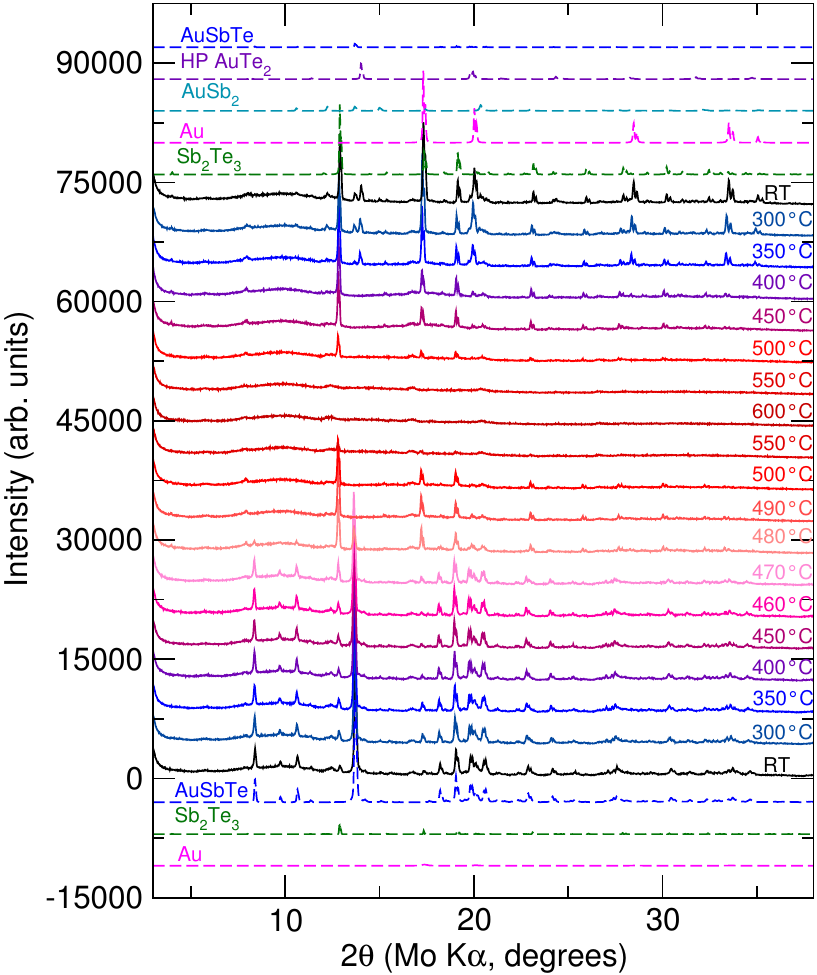}
    \caption{\textit{In situ} PXRD data collected on a crushed pellet of AuSbTe diluted with SiO$_2$ powder. The sample is sealed under vacuum in a quartz capillary and heated to 600\C before cooling back to room temperature. The pattern contributions of each phase (dotted lines) are shown for the room temperature scans before (bottom) and after (top) heating (see text for more details). }
    \label{fig:AuSbTeinsitu}
\end{figure}

Contributing to the difficulties associated with single crystal synthesis of AuSbTe is the fact that it is incongruently melting, which we have verified through \textit{in situ} PXRD and DTA measurements. Fig.\ \ref{fig:AuSbTeinsitu} shows PXRD scans collected at temperatures up to 600\C. The room temperature scan performed before heating is shown in black at the bottom of the figure. Below it, in dotted lines, are displayed the contributions of the AuSbTe (93.9(4)~wt.\%), Sb$_2$Te$_3$ (4.3(2)~wt.\%) and Au (1.8(3)~wt.\%) phases to the powder pattern. At 480\C, the signal of the AuSbTe phase diminishes while the signal of the Sb$_2$Te$_3$ phase becomes much stronger. Above 500\C, all signal is lost indicating that the sample is in the liquid phase. As the sample cools, Sb$_2$Te$_3$ precipitates. Below 400\C, Au and a phase that matches well to the high pressure diffraction pattern of AuTe$_2$\cite{Reithmayer:se0099} start to precipitate. Below 350\C, AuSb$_2$ and AuSbTe start to appear. The final room temperature scan is shown in black at the top of Fig.\ \ref{fig:AuSbTeinsitu}. Above it, in dotted lines, are displayed the contributions of the Sb$_2$Te$_3$ (47.9(3)~wt.\%), Au (28.2(3)~wt.\%), AuSb$_2$ (10.8(4)~wt.\%), high pressure AuTe$_2$ (10.5(2)~wt.\%) and AuSbTe (2.6(3)~wt.\%) phases to the powder pattern. A possible explanation for the presence of the AuTe$_2$ high pressure diffraction pattern at ambient conditions is that antimony substitutions are stabilizing the structure. No further investigation of this phase was performed. For simplicity, this phase will be referred to as HP-AuTe$_2$. Comparison of the PXRD patterns before and after the insertion of the furnace indicates that the broad peaks around 5.8$^{\circ}$, 7.9$^{\circ}$, 12.3$^{\circ}$ and 16.7$^{\circ}$ are due to scattering from the furnace. Consequently, these peaks still appear in the high temperature scans when the sample signal is absent due to its melting. 

When compared to the available literature, the phase transitions seen in the \textit{in situ} PXRD data differ from the ones predicted by the 67.1~at.\% Au, 32.9~at.\% Sb -- Sb$_2$Te$_3$ binary phase diagram\cite{Prince1990, ASMdiagram6732-4060}. Even if one ignores the temperature differences, which can be explained by insulation effects, the phases observed upon cooling are different. The binary phase diagram predicts that from the liquid, Sb$_2$Te$_3$ precipitates, then AuSbTe, then Au and more AuSbTe, followed by the solidification of Au, AuSb$_2$ and any remaining AuSbTe. The \textit{in situ} PXRD data suggests that from the liquid, Sb$_2$Te$_3$ precipitates, then Au and HP-AuTe$_2$, then AuSb$_2$ and a small amount of AuSbTe. An important difference concerns the resulting product; the \textit{in situ} PXRD shows that AuSbTe does not reappear at the temperature where it decomposed (near 480\C), and that other compounds such as HP-AuTe$_2$, which is not predicted by the binary phase diagram, are kinetically favorable. The phase diagram then overestimates the amount of AuSbTe reformed, which is minimal upon cooling of the \textit{in situ} PXRD sample.

\begin{figure}
    \centering
    \includegraphics[width=\columnwidth]{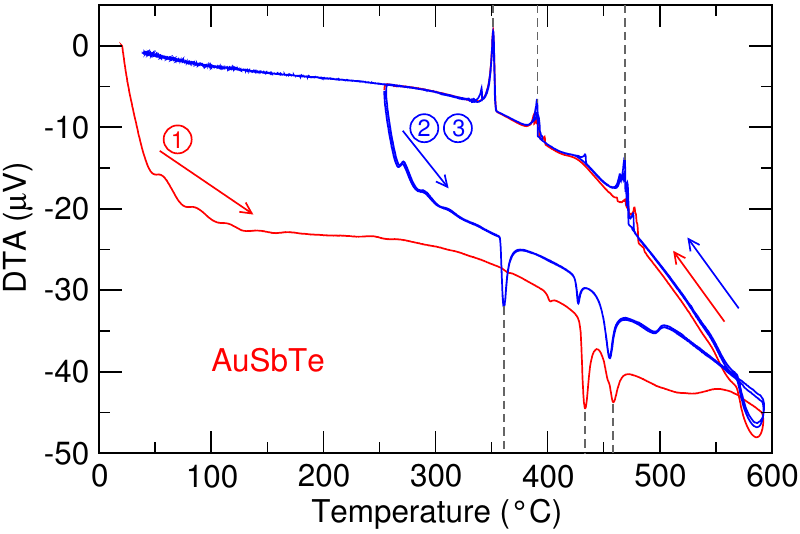}
    \caption{Differential thermal analysis measurements, defined as exothermic up, on a crushed pellet of AuSbTe. The first heating and cooling cycle is shown in red, while the subsequent cycles are shown in blue. Their starting point and direction are indicated by numbers and arrows. The grey dashed lines at 351\C (top), 361\C (bottom), 391\C (top), 434\C (bottom),  459\C (bottom) and 469\C (top) are a guide to the eye. }
    \label{fig:DTA_EAP0061}
\end{figure}

Three consecutive cycles of DTA (defined as exothermic up), shown in Fig.\ \ref{fig:DTA_EAP0061}, also confirm that AuSbTe is incongruently melting. The first DTA cycle, drawn in red, displays a strong trough around 434\C upon heating (bottom portion of the curve), which would correspond to the decomposition of the sample into Sb$_2$Te$_3$ and liquid. The next trough near 459\C would then indicate the transition towards the fully liquid phase. Upon cooling (top portion of the curve), the features around 500\C would be signs that Sb$_2$Te$_3$ precipitates. Based on the results of the \textit{in situ} PXRD, the peaks near 391\C and 351\C could correspond to the precipitation of Au and HP-AuTe$_2$, then AuSb$_2$ and a small amount of AuSbTe, which completes the re-solidification of the sample. Indeed, the fast heating and cooling rates of the DTA cycles should lead to reactions more similar to the step-and-hold temperature scans of the \textit{in situ} PXRD rather than the equilibrium reactions predicted by phase diagrams. The next DTA cycles, shown in blue, begin with different materials (Sb$_2$Te$_3$, Au, AuSb$_2$, AuSbTe and maybe HP-AuTe$_2$) compared to the first cycle, which had mostly AuSbTe. As such, the heating (bottom) portions of the curves display different features, like the trough around 361\C, which could correspond to the melting of AuSb$_2$ and some AuSbTe. The trough near 425\C is possibly Sb$_2$Te$_3$ melting or AuSbTe decomposing. The next heating features close to 455\C could then be associated with the melting of Sb$_2$Te$_3$, HP-AuTe$_2$ and/or the decomposition of AuSbTe. The broad feature close to 500\C could be the slow dissolution of gold. The cooling phases of all three cycles are then almost identical since they all start with the same liquid. 

The presence of different peaks in the second and third DTA cycles confirms the formation of additional phases and the incongruent melting of AuSbTe. It is possible that the first DTA cycle and the \textit{in situ} PXRD data show the same phase transitions, with the temperature differences being attributed to the different insulation, but it is also possible that the continuous cooling of the DTA experiment gave different products than the step-and-hold cool of the \textit{in situ} PXRD scans. Nevertheless, the DTA and \textit{in situ} PXRD measurements both confirm that AuSbTe is incongruently melting and that the kinetics of its formation are especially slow compared to the competing phases. This is in agreement with the fact that repeated grinding and annealing was necessary to obtain a synthetic sample of pampaloite.

\subsection{Electronic properties of AuSbTe}

\begin{figure}
    \begin{minipage}{\columnwidth}
        \raggedright
        (a) \\ \centering
        \includegraphics[width=\columnwidth]{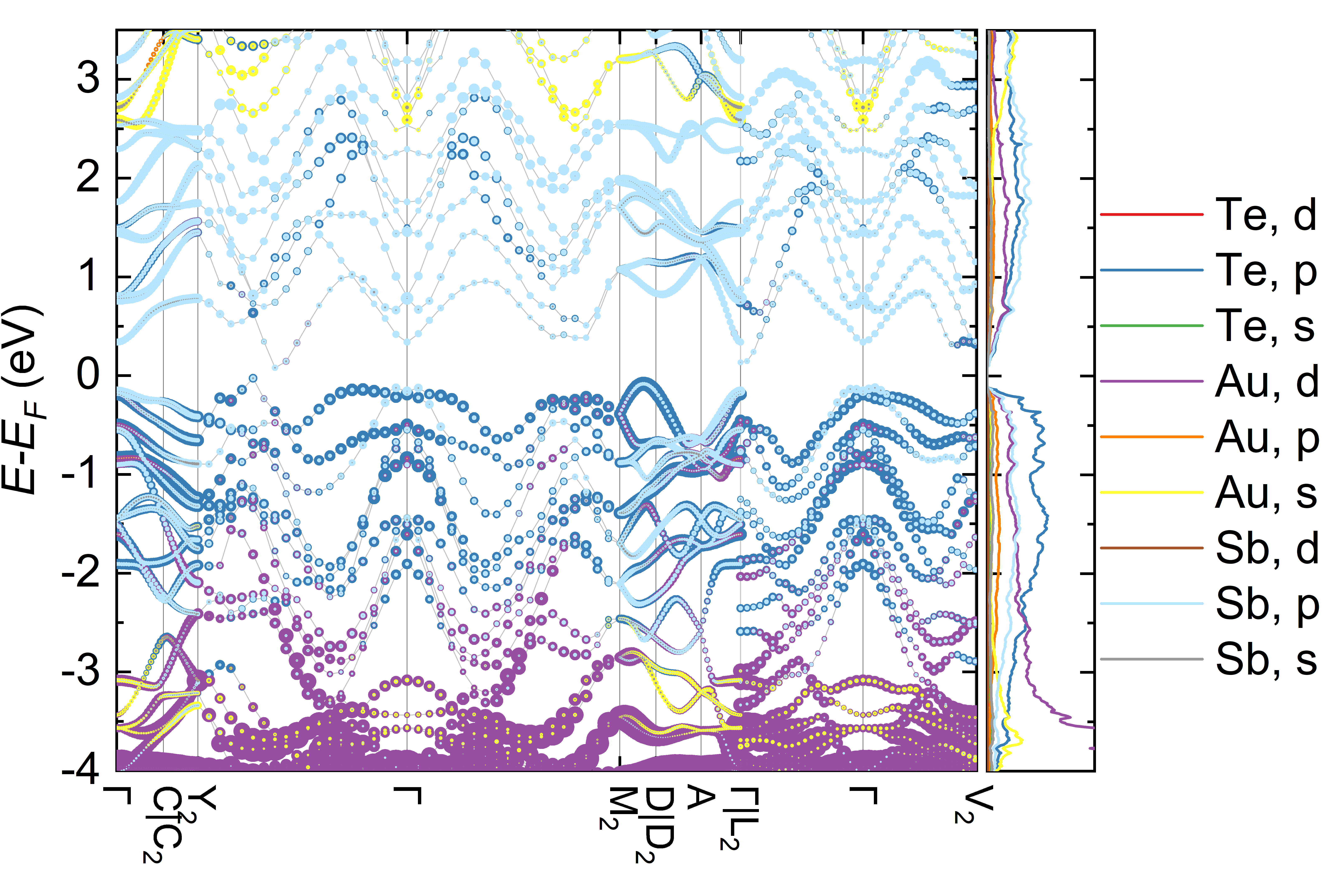}
    \end{minipage}
    \begin{minipage}{\columnwidth}
        \raggedright
        (b) \\ \centering
        \includegraphics[width=0.9\columnwidth]{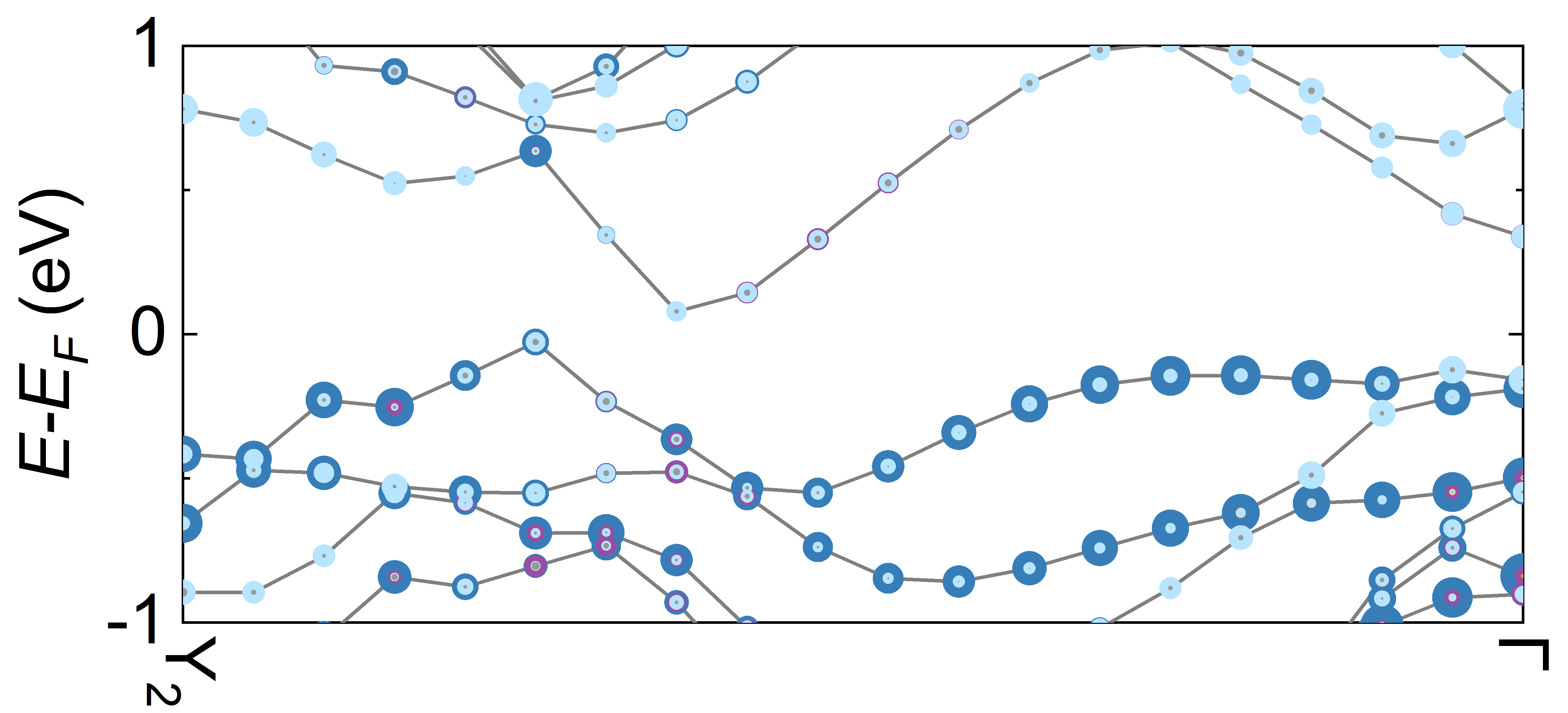}
    \end{minipage}
    \caption{(a) Band structure of AuSbTe (PBE, without SOC) with a valence band maximum and conduction band minimum both within the high-symmetry path Y$_2 \rightarrow \Gamma$. The indirect gap is 0.11~eV and the closest direct gap is 0.44~eV, both near Y$_2$. The weights on each band at a K-point indicate the contributions of electron density from each orbital. Overlapping weights signify hybridized orbitals. The density of states for each electron species is plotted on the right. (b) Zoomed-in view of the Y$_2 \rightarrow \Gamma$ high-symmetry path. }
    \label{fig:AuSbTe_DFT}
\end{figure}

We expect the heavy elements Au and Te to give rise to compounds with large spin-orbit coupling and narrow band gaps. The layered structure of AuSbTe could make it a cleavable and/or exfoliable semiconductor. The semiconducting behavior of AuSbTe was confirmed using our calculations, which revealed an indirect band gap of 0.11~eV near Y$_2$. The closest-approach direct-gap of the material also occurs near Y$_2$ and the direct gap at $\Gamma$ is 0.50~eV, see Fig.\ \ref{fig:AuSbTe_DFT}. The band structure shows the main contributions near the Fermi level are from the $p$-electrons of Te and Sb, with relatively minor contributions from the Au $d$-electrons. This is reminiscent of AuAgTe$_4$, where the formation of bonding-antibonding Te--Te dimers helps open up the band gap~\cite{Ushakov_2019}, unlike its sibling AuTe$_2$, which is a metal.\cite{Kudo2013, scAuTe2} 

The reported DFT calculations for PdSe$_{1.25}$Te$_{0.75}$ verbeekite, which crystallizes in the same space group as AuSbTe (C2/c), also show that it is an indirect band gap semiconductor.\cite{Wenhao2021} The proximity of the layers in AuSbTe and PdSe$_{2-x}$Te$_x$ verbeekite ($0.3 \leq  x \leq 0.8$)\cite{Wenhao2021} seems to create interlayer interactions strong enough to localize the electrons and open a gap in the band structure. Indeed, the unfilled Se, Te and Sb $p$-orbitals in the conduction bands and the similar shape of their contribution to the density of states are hallmarks of covalent Se--Se dimers in PdSe$_{2-x}$Te$_x$ verbeekite and covalent Sb--Te dimers in AuSbTe. Moreover, Bader charge analysis supports the notion that AuSbTe is covalently bonded, as the deviation from the number of valence electrons is at most $0.5$. The Bader charges for Au, Sb, and Te are 11.43, 6.04, and 4.52, respectively. Detailed analysis of the local environment for each Au atom is provided in the Supporting Information. The interlayer interactions in AuSbTe and PdSe$_{2-x}$Te$_x$ verbeekite contrast with the structurally analogous compound AuTe$_2$ in which an incommensurate modulation of the C2/m structure gives rise to various Te--Te interlayer bond distances. These variations in the electronic interactions could then leave charge carriers available for conduction, in agreement with the observed metallic behavior of AuTe$_2$.\cite{Kudo2013, scAuTe2} Upon chemical substitutions, one can obtain metallic low-temperature superconductors PdSe$_{2-x}$Te$_x$ (for $1.0 \leq x \leq 1.2$ and $x=2$)\cite{Wenhao2021} and Au$_{0.65}$Pt$_{0.35}$Te$_2$ in the P$\Bar{3}$m1 structure,\cite{Kudo2013}  which has larger interlayer distances. Supporting evidence can be found in the Supporting Information, where the disappearance of the band gap and the emergence of metallic behavior are observed upon enlarging the interlayer distances of AuSbTe. This modified structure prevents the potential formation of Te--Sb dimers and helps to shed light on their role in band gap origination.

\begin{figure}
    \centering
    \includegraphics[width=\columnwidth]{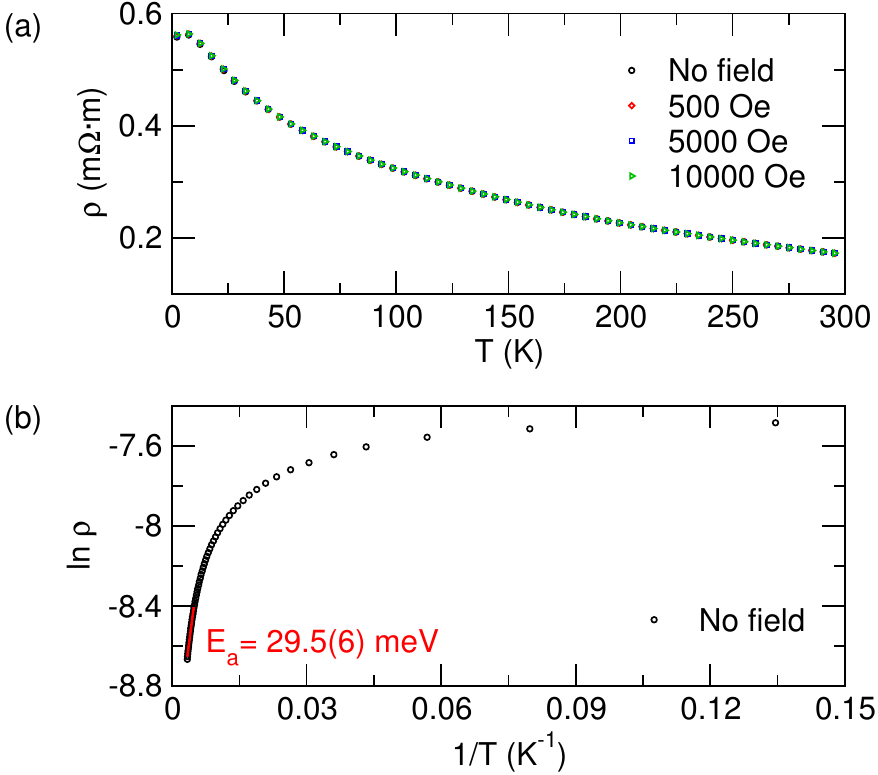}
    \caption{Temperature dependent four-point resistivity measurement of AuSbTe. The data is collected on heating. (a) Black circles, red diamonds, blue squares and green triangles show data collected without an applied magnetic field, with 500~Oe, 5000~Oe and 10000~Oe, respectively. The average error coming from the instrument and geometry is on the order of 7\%. (b) The Arrhenius plot displays a linear behavior only at high temperatures, which is expected considering the minor (less than or equal to 8~wt.\%) metal impurities present in the sample. The activation energy is being fitted for between 200 and 300~K (red). }
    \label{fig:AuSbTe_RvsT}
\end{figure}

To confirm the semiconducting behavior of AuSbTe, we performed four-point resistivity measurements. Fig.\ \ref{fig:AuSbTe_RvsT}(a) shows the temperature dependent resistivity with various applied magnetic fields. The overlap of the data points with and without magnetic fields indicate negligible magnetoresistance, which is confirmed by the low temperature field sweep found in the Supporting Information. The global increase of the resistivity as the temperature decreases indicates a dominant semiconducting behavior associated with the main AuSbTe phase. At low temperature, a small decrease in the resistivity suggests that some minor metallic impurities may contribute to the signal. The presence of small amounts (less than or equal to 8~wt.\% combined) of Au and Sb$_2$Te$_3$ in the resistivity sample was confirmed by SEM-EDS analysis (see the Supporting Information). These impurities are known to be metallic\cite{Sultana_2018} and explain this small decrease in resistivity at low temperature and the non-linear shape of the Arrhenius plot seen in Fig.\ \ref{fig:AuSbTe_RvsT}(b). No other phases were found in the sample, indicating that the main AuSbTe phase must be the source of the semiconducting behavior seen in the data. The shape of the temperature-dependent resistivity reflects this mixing between the dominant semiconducting AuSbTe and the minor metallic impurities. As such, the Arrhenius plot displays an almost linear shape at high temperatures. Linear fits between 50 and 100~K (not shown in Fig.\ \ref{fig:AuSbTe_RvsT}(b)) and between 200 and 300~K (shown in red in Fig.\ \ref{fig:AuSbTe_RvsT}(b)) yield activation energies of 4.3(2)~meV and 29.5(6)~meV, respectively. The high temperature value of 29.5(6)~meV is a better estimate of the activation energy of AuSbTe. Indeed, upon cooling, the resistivity of AuSbTe increases, while the resistivity of the metallic impurities decreases, allowing more transport through the impurities, which lowers the activation energy. Temperature dependent Van Der Pauw measurements (see Supporting Information) displayed the same shape as in Fig.\ \ref{fig:AuSbTe_RvsT}, confirming that the observed behavior is directly caused by the sample and not by some experimental artifact.

\subsection{Structure of \aattt}

\begin{figure}
    \centering
    \includegraphics[width=0.95\columnwidth]{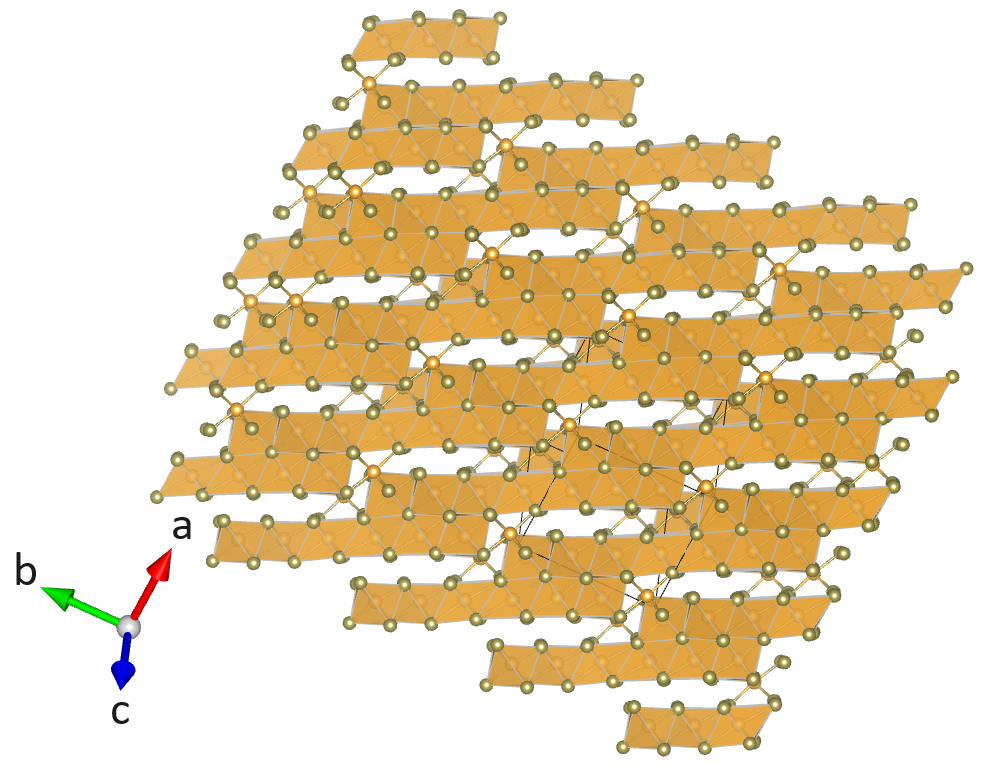}
    \caption{Crystal structure of montbrayite, showing only the Au (gold) and Te (green) atoms for simplicity. Te atoms form distorted octahedra (shown in gold) around the Au atoms, creating displaced double layers reminiscent of the Bi$_2$Se$_3$ structure.}
    \label{fig:Au2Te3_structure}
\end{figure}

Montbrayite \aattt crystallizes in the low symmetry structure P$\Bar{1}$. Fig.\ \ref{fig:Au2Te3_structure} shows that the tellurium ions (green) arrange themselves into distorted octahedra (gold polygon) around the gold ions (the partial occupation of antimony is omitted in the figure for clarity). The Au--Te bonds that form the [AuTe$_6$] polyhedra are all between 2.66 and 3.33~\r{A}.\cite{Bindi2018} These lengths are very similar to those of the Au--Te and Au--Sb bonds forming the [AuSb$_3$Te$_3$] octahedra in AuSbTe, but unlike the flat single layers found in pampaloite, the montbrayite polyhedra form double layers with periodic displacements that give them a staircase appearance. The double layers make the montbrayite structure more analogous to that of Bi$_2$Se$_3$. The montbrayite layers are connected by gold ions that only make contacts with five tellurium ions. The sixth bond is with another gold ion and has a length of 3.154(2)~\r{A}.\cite{Bindi2018} Between the layers, the shortest Te--Te distance is 2.83~\r{A},\cite{Bindi2018} the same length as the Sb--Te interlayer bond in AuSbTe. 

\begin{figure}
    \centering
    \includegraphics[width=\columnwidth]{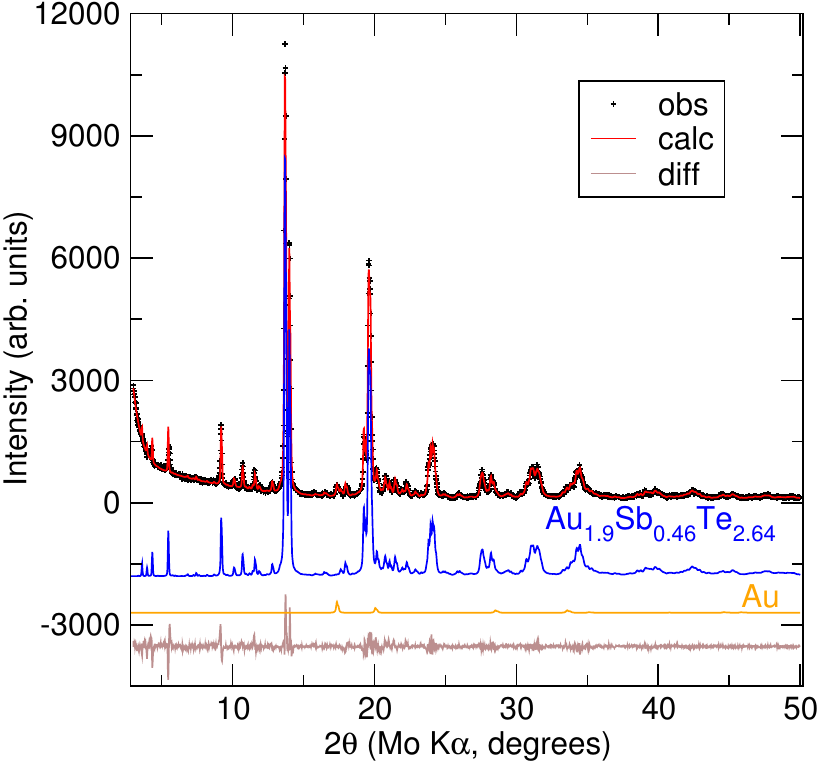}
    \caption{Rietveld refinement of the room temperature PXRD data of the \aattt pellet used for \textit{in situ} PXRD and DTA measurements. The main phase is fitted to the reported structure of Au$_{1.81}$Sb$_{0.11}$Bi$_{0.12}$Te$_{3.04}$\cite{Bindi2018} even though the chemical composition of the sample is slightly different. A minor (1.2(1)~wt.\%) gold impurity is present in the data. }
    \label{fig:EAP0030_XRD}
\end{figure}

The PXRD patterns for the \aattt samples used in this study were fitted to the reported structure for Au$_{1.81}$Sb$_{0.11}$Bi$_{0.12}$Te$_{3.04}$.\cite{Bindi2018} For the data shown in Fig.\ \ref{fig:EAP0030_XRD}, the lattice parameters were refined to be $a$ = 10.811(3)~\r{A}, $b$ = 12.102(4)~\r{A}, $c$ = 13.369(6)~\r{A}, $\alpha$ = 107.84(1)$^{\circ}$, $\beta$~= 104.58(1)$^{\circ}$, and $\gamma$ = 97.530(7)$^{\circ}$. The atom positions were fixed and the overall weighted profile R-factor and goodness of fit are 10.74\% and 2.34, respectively. Despite the slightly different chemical composition, we find that the P$\Bar{1}$ structure of Au$_{1.81}$Sb$_{0.11}$Bi$_{0.12}$Te$_{3.04}$ better fits our data compared to the previously reported P1 structure for Au$_2$Te$_3$.\cite{Bachechi1971} A few small angle peaks in the diffraction pattern of the non-centrosymmetric P1 space group were absent from our patterns. Earlier refinements of PXRD data for \aattt also reported some deviations between the sample's diffraction pattern and the P1 structure.\cite{Nakamura2002} As in this previous work, we conclude that the initial indexing of the montbrayite phase was not yet properly solved. The more recent indexing with the centrosymmetric P$\Bar{1}$ space group does not have any additional peaks or large deviations from our powder patterns and addressed the structural concerns associated with the P1 structure.\cite{Bindi2018} 

Fig.\ \ref{fig:EAP0030_XRD} is representative of the PXRD pattern refinement results for all the samples, except for slightly different impurity contents. The sample used for \textit{in situ} PXRD and DTA (seen in Fig.\ \ref{fig:EAP0030_XRD}) has 1.2(1)~wt.\% of gold, the one used for resistivity measurements has 0.6(1)~wt.\% of gold (see the Supporting Information), and the sample used for STEM imaging has 6.5(5)~wt.\% of AuTe$_2$ and 1.6(3)~wt.\% of Sb$_2$Te$_3$ (see the Supporting Information). No other phases were observed in the PXRD patterns. SEM-EDS confirmed the presence of a minimal gold impurity in the sample used for \textit{in situ} PXRD and DTA and revealed small Te-rich regions in all samples. The specimens appear otherwise homogeneous. The SEM-EDS results are available in the Supporting Information.

\begin{figure}
    \centering
    \includegraphics[width=\columnwidth]{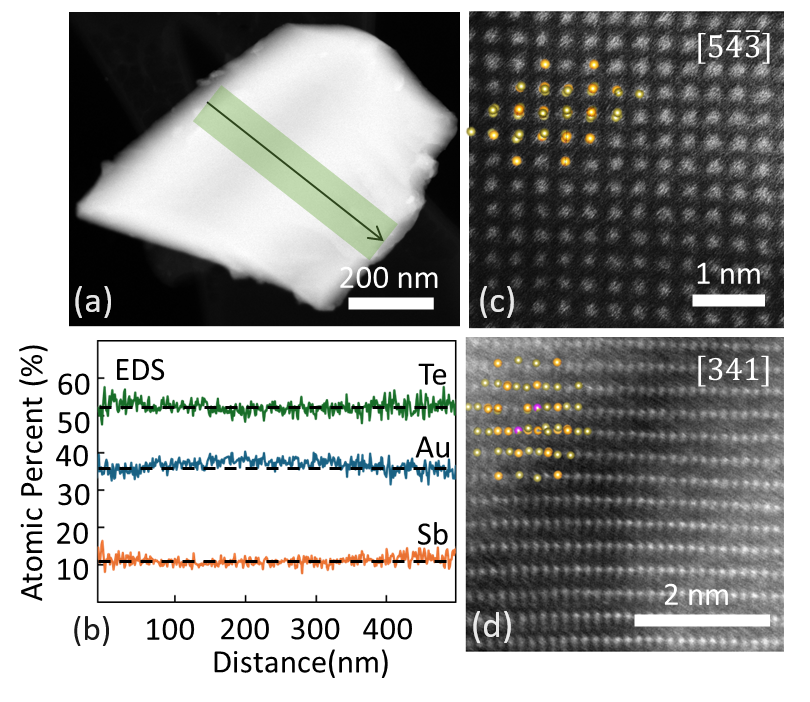}
    \caption{(a) Scanning transmission electron microscopy images of an isolated \aattt particle used for EDS analysis (taken from marked region) and High-Resolution STEM imaging. (b) EDS Spectra for Au, Te, and Sb across the particle of interest with overall average marked(black dotted line). (c) HR-STEM images of the particle edge structure near the [5$\Bar{4}$$\Bar{3}$] orientation, with structural overlay in the top left corner. (d) HR-STEM images of the particle edge structure near the [341] orientation, with structural overlay in the top left corner.}
    \label{fig:Au2Te3-TEM}
\end{figure}

Scanning transmission electron microscopy (STEM) was used to confirm the chemical composition and structure of our compound. Fig. \ref{fig:Au2Te3-TEM} shows STEM images and EDS of the \aattt crystal. EDS profiles of Te, Au, and Sb are displayed for the shown particle over the marked region. Taken as atomic percent, Te is found to maintain 52.24 $\pm$ 1.62\%, Au maintains 36.64 $\pm$ 1.59\%, and Sb maintains 11.12 $\pm$ 1.24\%. This is in reasonably good agreement with the \aattt formula, which corresponds to 52.8~at.\% Te, 38~at.\% Au and 9.2~at.\% Sb. Edges of a 500~nm particle were used for high resolution imaging with atom spacing matching the [5$\Bar{4}$$\Bar{3}$] and [341] orientations. Atom spacing for the [5$\Bar{4}$$\Bar{3}$] orientation is 2.88~Å matching the expected 2.89~Å projected distance for the XRD derived model\cite{Bindi2018} as shown with the atomic model overlay. The [341] has atomic spacings of 1.14~Å by 2.494~Å compared to the expected projected distances of 1.14~Å by 2.441~Å, indicating approximate agreement with the expected atomic structure.

\subsection{Stability of \aattt by synthesis, diffraction and calorimetry}

Differentiating \aattt from other gold telluride phases can be subtle. In particular, the powder x-ray diffraction pattern of \aattt is very similar to the diffraction pattern of AuTe$_2$. A diffraction peak near 4.42~\r{A} \textit{d}-spacing ($\sim$ 9.19$^{\circ}$ in Fig.\ \ref{fig:EAP0030_XRD}), and other low angle peaks, present in \aattt but not in AuTe$_2$, and a diffraction peak near 5.02~\r{A} \textit{d}-spacing (which would be at $\sim$ 8.07$^{\circ}$ in Fig.\ \ref{fig:EAP0030_XRD}) present in AuTe$_2$ but not in \aattt, can help to distinguish the phases. These differences helped to confirm the results of synthesis attempts with lower antimony contents. Indeed, the PXRD patterns of samples with compositions Au$_2$Te$_3$ and Au$_{1.89}$Sb$_{0.11}$Te$_3$ lacked any indication of the montbrayite phase, while the composition Au$_{1.89}$Sb$_{0.23}$Te$_{2.88}$ gave a small peak near 9$^{\circ}$ associated with montbrayite. Unlike \aattt, the three antimony-deficient compounds had significant AuTe$_2$ and Au impurities, suggesting that these compositions do not result in stable montbrayite. This is in agreement with Nakamura and Ikeda's conclusion that, within the Au--Sb--Te phase space, the stable composition of montbrayite is \aattt with a very narrow region of stability.\cite{Nakamura2002} As previously reported,\cite{PeacockThompson1946, Markham1960, Cabri1965} our own results confirm that the stoichiometric formula Au$_2$Te$_3$ is unstable to AuTe$_2$ and Au. 

\begin{figure}
    \centering
    \includegraphics[width=\columnwidth]{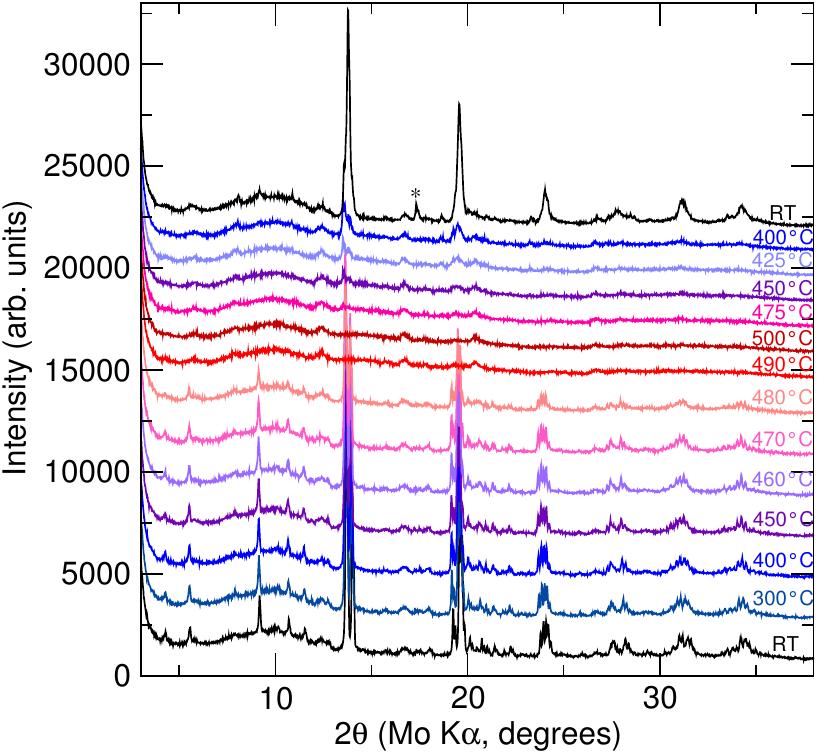}
    \caption{\textit{In situ} PXRD data collected on a crushed pellet of \aattt diluted with SiO$_2$ powder. The sample is sealed under vacuum in a quartz capillary and heated to 500\C before cooling back to room temperature. A small gold diffraction peak is labeled by a star above the final room temperature scan (see text for more details). }
    \label{fig:Au2Te3insitu}
\end{figure}

Our stability investigations of \aattt revealed that it is congruently melting, in contrast to the more complicated phase transitions found for AuSbTe. Its \textit{in situ} PXRD data is displayed in Fig.\ \ref{fig:Au2Te3insitu}. The lower intensity of the data at 480\C, which is particularly noticeable due to the decrease in size of the peak close to 9$^{\circ}$, is indicative of melting past 470\C. At 490\C, all the major peaks are suppressed, suggesting that the sample is completely in the liquid phase. The peak intensities slowly increase as the temperature decreases, with the final room temperature data being misshaped due to the fact that the sample is slow cooled. The final scan shows that the sample is still mostly composed of \aattt with minimal decomposition into competing phases AuTe$_2$ (10.3(6)~wt.\%) and Au (3.4(2)~wt.\%), with a diffraction peak of the latter being indicated by a star in Fig.\ \ref{fig:Au2Te3insitu}. Refinement of the final room temperature data is available in the Supporting Information. Comparison of the PXRD patterns before and after the insertion of the furnace indicates that the broad peaks around 5.8$^{\circ}$, 7.8$^{\circ}$, 12.3$^{\circ}$ and 16.6$^{\circ}$ are due to scattering from the furnace. Consequently, some of these peaks, such as the ones at 12.3$^{\circ}$ and 16.6$^{\circ}$, still appear in the high temperature scans when the sample signal is absent due to its melting. 

\begin{figure}
    \centering
    \includegraphics[width=\columnwidth]{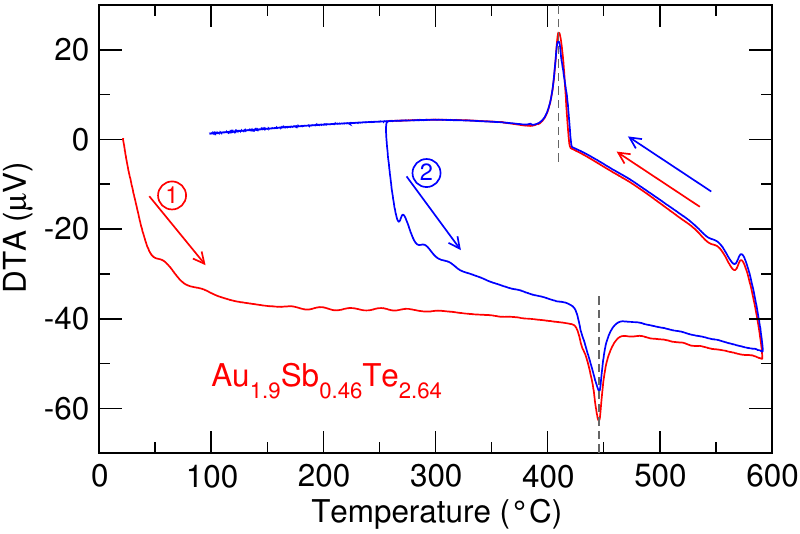}
    \caption{Differential thermal analysis measurements, defined as exothermic up, on a crushed pellet of \aattt. The first heating and cooling cycle is shown in red, while the second cycle is shown in blue. Their starting point and direction are indicated by numbers and arrows. The grey dashed lines at 410\C and 446\C are a guide to the eye. }
    \label{fig:DTA_EAP0030}
\end{figure}

The stability of \aattt is also confirmed by two consecutive cycles of DTA (defined as exothermic up) shown in Fig.\ \ref{fig:DTA_EAP0030}. Upon heating, troughs at 446\C indicate that the sample is melting. Upon cooling, peaks at 410\C indicate that the sample solidifies. The overlap of the troughs and peaks indicate the repeatability of this heating and cooling cycle, while the absence of additional strong features suggest that no other transition is taking place. The solidification temperature of \aattt at 410\C is akin to the 410\C melting temperature of Au$_{1.89}$(Sb,Pb)$_{0.11}$Te$_{2.88}$Bi$_{0.12}$ found by Bachechi,\cite{Bachechi1972} although the synthetic sample used in that study has a reported composition that we did not attempt to replicate. 

The \textit{in situ} PXRD data and the DTA measurements show that synthetic \aattt is congruently melting. Samples of \aattt annealed above the melting point, at 450\C, 500\C and 600\C for two days, then water quenched, showed signs of melting and decomposition. PXRD analysis indicated the presence of 47.8(7), 44.7(6) and 27.9(5)~wt.\% of AuTe$_2$, the rest of the samples being mostly \aattt with small amounts of gold. This goes to show that quenching is sufficiently fast to prevent the compound from fully reaching its equilibrium structure. This is in agreement with previous reports on natural and synthetic compounds which found that montbrayite decomposed into AuTe$_2$ and gold at high temperatures.\cite{PeacockThompson1946, Bachechi1972, BlachnikGather1976}

\subsection{Metallicity of \aattt}

\begin{figure}
    \centering
    \vspace{-8mm}
    \includegraphics[width=\columnwidth]{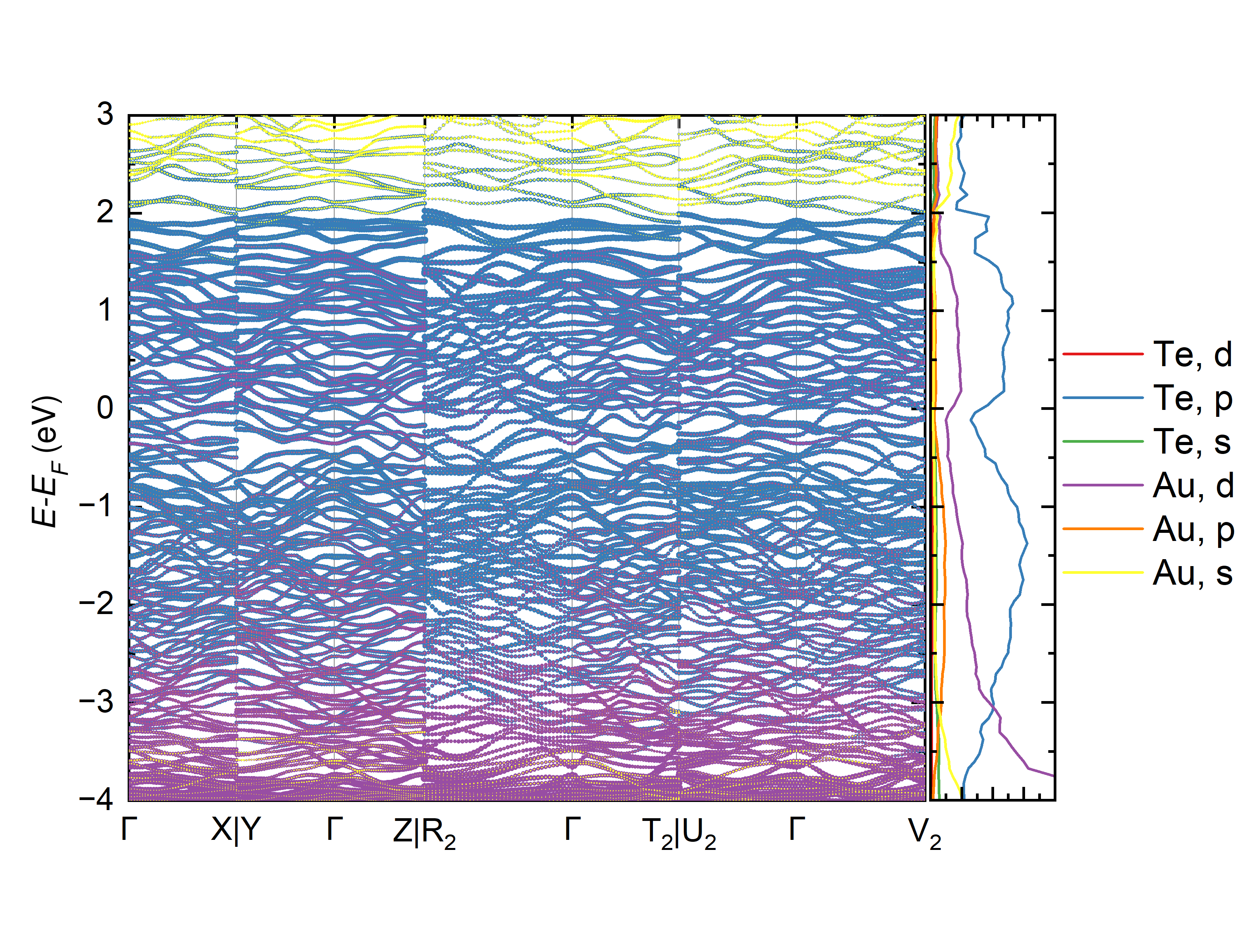}
    \vspace{-8mm}
    \caption{The DFT-PBE band structure of an idealized Au$_2$Te$_3$ cell, with no Sb substitution, is highly metallic with many Au $d$ and Te $p$ states at the Fermi energy. The density of states for each electron species is plotted on the right. }
    \label{fig:Au2Te3_DFT}
\end{figure}

Density functional theory (DFT) calculations of the band structure and density of states of Au$_2$Te$_3$ show it is metallic (see Fig.\ \ref{fig:Au2Te3_DFT}). The large number of bands is a hallmark of the large unit cell and low $P\overline{1}$ symmetry of Au$_2$Te$_3$.
The ideal chemical formula of montbrayite, Au$_2$Te$_3$, was used for the calculations to reduce the computational complexity. This composition could lead one to assume that the compound is made up of Au$^{3+}$ and Te$^{2-}$ ions, but the similar shape of the density of states of the gold and tellurium atoms indicate that the compound is covalent, not ionic. The presence of covalent bonds is consistent with the metallic behavior, the partial substitution of gold and tellurium with antimony necessary to stabilize the compound, and the close proximity of these elements in the periodic table. Bader charge analysis also supports the notion that the compound is covalently bonded, as the deviation from the number of valence electrons is less than $0.5$. More details about the Bader charge analysis are provided in the Supporting Information, along with an enlarged view of the energy region around the Fermi level in Fig.\ \ref{fig:Au2Te3_DFT}. 

\begin{figure}
    \centering
    \includegraphics[width=\columnwidth]{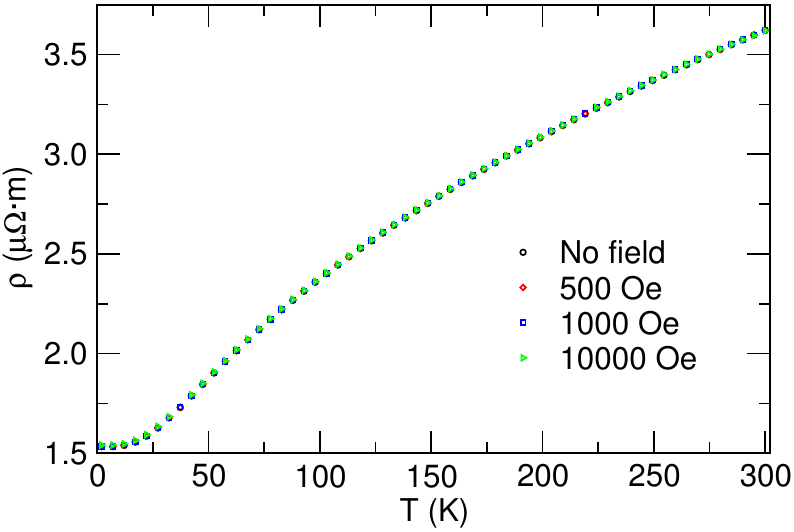}
    \caption{Temperature dependent four-point resistivity measurement of \aattt. The data is collected on heating and gives a residual resistivity ratio of 2.36. Black circles, red diamonds, blue squares and green triangles show data collected without an applied magnetic field, with 500 Oe, 1000 Oe and 10000 Oe, respectively. The average error coming from the instrument and geometry is on the order of 10\%.}
    \label{fig:Au2Te3_RvsT}
\end{figure}

The metallic nature of \aattt was confirmed using four-point resistivity measurements. Fig.~\ref{fig:Au2Te3_RvsT} shows a decrease of the resistivity as the temperature decreases. The residual resistivity ratio is 2.36. Data collected with applied magnetic fields of 500~Oe, 1000~Oe and 10000~Oe almost overlap the data collected without a magnetic field, indicating that no significant magnetoresistance is observed. This is confirmed by the field sweep at $T=2$~K found in the Supporting Information. Negligible magnetoresistance has been reported in the similar metallic gold tellurides AuTe$_2$ and Au$_{0.65}$Pt$_{0.35}$Te$_2$.\cite{Millard1969, Kudo2013} The 3.6~$\mu\Omega\cdot$m room temperature resistivity of \aattt is also similar to that of AuTe$_2$.\cite{Kudo2013} 

\section{Conclusions}

We have presented a comprehensive investigation of the electronic properties of pampaloite and montbrayite. Despite previous reports, their band structure and resistivity were still unknown and information about their thermal and chemical stability was at times conflicting. We find both compounds to be compositionally well-defined and stable. The narrow band gap of AuSbTe is intriguing, in combination with its layered structure which could indicate a cleavable and/or exfoliable semiconductor. 

Four-point resistivity measurements revealed the semiconducting behavior of AuSbTe and the metallic nature of \aattt, in agreement with our DFT-calculated band diagrams and density of states. The DFT-PBE method gave a calculated indirect band gap of 0.11~eV for AuSbTe, which is likely an underestimate.

We investigated the thermal and chemical stability of AuSbTe and \aattt by reporting the effectiveness of various syntheses and by performing DTA and \textit{in situ} PXRD experiments. Our results showed that AuSbTe is incongruently melting and that re-grinding and re-annealing the compound multiple times is essential to obtain a high quality polycrystalline sample. We confirmed the narrow stable chemical composition of \aattt and demonstrated that it is congruently melting with a few possible heating profiles resulting in a mostly pure polycrystalline product. 

Overall, our exploration of the properties and syntheses of gold antimony tellurides clarifies inconsistencies regarding synthetic montbrayite and emphasized the potential for AuSbTe to be a layered semiconductor. Both compounds remain to be synthesized as large single crystals, but the results of our numerous syntheses will be useful in guiding future attempts. 

\section*{Supporting Information}

Additional synthesis details, powder XRD refinements, SEM-EDS images, resistivity measurements, band structures, and Bader charge analysis. (PDF)

\begin{acknowledgments}
Materials syntheses, transport, and microstructure characterization were supported by the Center for Quantum Sensing and Quantum Materials, an Energy Frontier Research Center funded by the U. S. Department of Energy, Office of Science, Basic Energy Sciences under Award DE-SC0021238. The authors acknowledge the use of microscopy facilities at the Materials Research Laboratory Central Research Facilities, University of Illinois, and the use of facilities and instrumentation supported by NSF through the University of Illinois Materials Research Science and Engineering Center DMR-1720633. Computations conducted by RZ and CP were supported by the U. S. Department of Energy, Office of Science, Basic Energy Sciences under Award DE-AC02-76SF00515. This research used resources of the National Energy Research Scientific Computing Center, a DOE Office of Science User Facility supported by the Office of Science of the U.~S. Department of Energy under Contract No. DE-AC02-05CH11231 using NERSC award BES-ERCAP0027203. RZ and CP thank Brian Moritz and Chunjing Jia for insightful discussions.

\end{acknowledgments}

\bibliographystyle{unsrt}

\bibliography{Au-Sb-Te.bib}

\end{document}